# Student Resources in quantum mechanics, or why students need meta-resources.

*Keith Oliver, The Ohio State University, oliver@mps.ohio-state.edu*
*Lei Bao, The Ohio State University, lbao@mps.ohio-state.edu*

We are trying to identify resources students use to reason in quantum mechanics. In this process we realize that a student must have not only the right conceptual resources available but also sophisticated resources for evaluating and controlling their thought processes. We will discuss examples from student interviews to illustrate our point.

**Resources**

Hammer[1] has given us a useful way of applying our knowledge of cognitive structure to research and teaching. He suggests we can gain insight for research and teaching by looking at students' cognitive structures as resources for learning. He compares student learning to programming in a computer language. A programmer does not have to start from scratch in machine language to write a program. There are many resources she may use. At the level of the operating language some procedures and methods are defined. In the programming language a large base of useful functions are available for the programmer and a programmer will have available a large library of pre-assembled subroutines and procedures that he can insert to their program as needed. For the programmer these resources are neither right nor wrong out of the context where they are applied; each is useful in a particular domain. Errors arise from applying the resource in a domain where it is not well suited for the task.

Similarly students bring to class a large library of resources for thinking. Some of the things they bring are part of their cognitive "operating system", like the way memory is stored and retrieved. Since this is always on, it is not a resource for student learning. It may still be useful for a teacher to understand but since it can not be changed or turned off we will not call it a resource. Other cognitive structures like *raw intuitions*[2] or *primitives*[3] are basic operational procedures of the programming language. These are resources. Larger structures like *mental models*[4] are also resources. Physics education researchers have done much to identify the cognitive structures students bring with them to physics class. Some like Bao[4] and diSessa[5] have examined details of the structures on a finer grain. We can improve research and instruction by viewing most of the cognitive structures students bring with them as resources for learning. Calling these structures *resources* brings up familiar cognitive structures in our own minds that help us understand how this knowledge of student thinking can help us teach.

**Resources seen in quantum mechanics**

In the quantum mechanics classroom there is a rich and interesting variety of student thinking. We have begun research to look at the resources students use in understanding quantum mechanics.

Two sets of resources from classical mechanics are needed for learning in quantum mechanics: the students' ideas about waves and particles, including a number of components and sub-ideas. Students may or may not be good at using the ideas and may not have coherent ideas about waves or particles but these ideas are discussed from the start of nearly every quantum mechanics curriculum.

The wave ideas are resources brought in to explain the way quantum systems propagate. The particle ideas are resources brought in to explain how quantum systems interact with measurement devices. This is the famous wave-particle duality. Students and experts both struggle with the idea of wave particle duality. Most students have not been asked to merge the two cognitive structures (developed in different contexts) into one new structure.

**Other resources in quantum mechanics**

Physics education researchers have observed students using a variety of resources from outside physics contexts to negotiate the process of constructing the new cognitive structure. For example Rosenberg[6] has seen students making reference to interpersonal relations in explaining linear combinations of energy eigenstates. He has shown students treating the wave function collapse onto one energy eigenvalue as a person deciding between two options, such as what to have for dinner. He has seen similar behavior in a number of students at various levels, from sophomore physics majors to first year graduate students. Feynman[7] in his famous "Lectures on Physics", explains that the wave function "smells out" all possible paths with its wavelength. He uses the idea of a dog, sniffing out the paths an animal takes to help students make sense of the quantum behavior.

Feynman did not advocate a theory of quantum mechanics based on intelligent particles. Rosenberg suggested that the students in his study are using the *deciding* resource from personal relationships to de-emphasize a strong, in this case unproductive cognitive element of *object permanence*. Piaget[8] suggests that *object permanence*, the expectation that objects continue to exist when not directly observed, develops early in children and forms a foundation for much of our thought processes throughout our lives. This is almost always a good thing. There are only a few exceptions, for example, a quantum system or the state of a person's mind.

**The danger of other resources**

The students in Rosenberg's study used *deciding* as a resource effectively. They used it to distract object permanence from the property of energy. There is a large danger here for students. If they attach too much significance this resource of deciding it will interfere with their reasoning. For example deciding may not be productive for students in thinking about multi-particle systems. (Do the many particles make up their collective mind?)

In our own studies we asked students to reason about a system in a linear combination of energy eigenstates. Todd is a junior physics major; the following discussion took place several weeks into his quantum mechanics course.

*"Well, say that you are in your house, there is some percent chance that you are going to be in the bedroom or the bathroom or the kitchen or the living room… So you could make a function of the probability and if someone who can't see into your house (pause) No, that is not exactly right because you are in one of those rooms as opposed to the energy which is, um (pause)*

*Shoot, maybe it is more like [the particle has and energy and we just don't know what it is]"*

During the course of our discussions, Todd indicated that he felt that comparing quantum mechanical systems to social systems was appropriate. In this excerpt you can see how his analogy to a person failed; he recognized the conflict; he evaluated it and chose to follow the analogy instead of what he thought the answer should be from quantum mechanics. Later in the interview, he recognized the problem again and chose to stick with what he remembered from class, but he still could not resolve the difference between what he expected from his personal analogy and what he remembered from class.

**Meta-resources**

There is a cognitive minefield that students need to negotiate as they try to construct an understanding of quantum mechanics from the resources they possess coming into class. Some instructors suggest that instruction should avoid any references to problematic resources[9]. To do this they ask the students to turn off all their prior ideas and learn new ones. Continuing the computer analogy, this like is the instructor (as the programmer) trying to use the resources to program the students' brains for them. Rosenberg[10] suggests that many of these resources, such as object permanence, cannot be turned off, and that students will continue to try to make sense of new ideas using their available resources outside of class regardless of how we try to tell them to think. So programming for them may not be possible.

By contrast, Elby[2] when teaching introductory physics, tries to teach his students to evaluate their use of resources and control their own learning. In the computer analogy, this is like asking the students to be programmers in their own brains. It seems in the long run that teaching the students how to learn for themselves will the more effective objective.

Let's call the resources students can use to evaluate and control their own thought processes *meta-resources*. Meta-resources include things like metacognition, epistemology, affect and expectations.

**Examples of Meta-resources**

At this point one could ask, does one need a new set of resources to learn quantum mechanics, in addition to conceptual resources? Simply put, no; they use what they have when they come into class. What we should ask is what do they have when they come in. We have one example already. Todd, using his analogy to personal relations, noticed a conflict between his class knowledge and his analogy. This shows that he was aware of his thoughts and was comparing them to his memory of class discussion or textbook explanation. This is a form of metacognition, being aware of ones own thoughts. Next, he recognized that the two threads had different outcomes. We can view this metacognition as a resource for monitoring the use of other resources. He expected the two threads to lead to the same answer. He then has to evaluate the two ideas. It seems he may expect that knowledge in physics should be coherent, that is, different paths should lead to compatible ends and ideas from different areas should not contradict each other. This type of belief has been studied by Hammer[11] in introductory physics students. We can view these beliefs as resources for students to use in evaluating other resources.

After Todd notices the conflict, he tries to resolve it. He has to decide which is more important, his recollection of class or his analogy to personal systems. He chose to stay with his analogy. Another possible course of action would be to look for a third way of thinking and comparing the three. These are more examples of meta-resources.

One would then ask if a student, coming into our quantum mechanics classroom, has these meta-resources available. We think the answer is yes, but not as well developed as they might be. Consider Rich, a sophomore physics major taking a modern physics class focused mainly on quantum mechanics. In an interview we were discussing using a semi-classical model for the photoelectric effect and a traditional quantum model of the same effect. In the interview he felt they both gave the same predictions in this immediate context. He was then prompted to explain how to choose one model over the other. His reply shows some of the resources he used to answer.

> *"Well if you just want to get the answer in this situation you could think about it either way but it is the deeper understanding of what is going on that could benefit from understanding what is really going on…"*

He has some idea of what to expect from a physics model, that it should be applicable in many different situations. In the discussion he indicated he expected the semi-classical model we discussed to fail in other situations. His reasoning is not fully developed, apparent by the circular nature of his comment. But the resources are there ready to be developed.

## Conclusions

Students have many resources available to learn quantum mechanics. Some of the conceptual resources are from their physics knowledge and some come from other areas. In order for them to do well they need meta-resources for monitoring, controlling and judging their thought processes. Their conceptual and meta-resources are often underdeveloped.